\documentclass[11pt,amsmath,amssymb,nofootinbib]{revtex4}
\usepackage{amsmath}
\usepackage{dcolumn}
\usepackage{bm}
\usepackage[all, knot]{xy}
\xyoption{arc}
\begin{document}
\title{On the 5D differential calculus and translation transformations\\
 in 4D $\kappa$-Minkowski noncommutative spacetime}
\author{Giovanni Amelino-Camelia}
\email{giovanni.amelino-camelia@roma1.infn.it}
\affiliation{Dipartimento di Fisica \\
Universit\`a degi Studi di Roma ``La Sapienza"\\
and Sez.~Roma1 INFN\\
P.le Moro 2, Roma 00185, Italy}
\author{Antonino Marcian\`o}
\email{antonino.marciano@roma1.infn.it}
\affiliation{Dipartimento di Fisica \\
Universit\`a degi Studi di Roma ``La Sapienza"\\
and Sez.~Roma1 INFN\\
P.le Moro 2, Roma 00185, Italy}

\author{Daniele Pranzetti}
\email{danielepra@libero.it}
\affiliation{Dipartimento di Fisica\\
Universit\`a degi Studi di Roma ``Roma Tre"\\
via della Vasca Navale 84, Roma 00146, Italy }

\begin{abstract}
\begin{center}
{\bf Abstract}
\end{center}
\noindent \noindent We perform a Noether analysis for a description
of translation transformations in 4D $\kappa$-Minkowski
noncommutative spacetime which is based on the structure of a 5D
differential calculus. The techniques that some of us had
previously developed (hep-th/0607221) for a description of
translation transformations based on a 4D differential calculus
turn out to be applicable without any modification, and they allow
us to show that the basis usually adopted for the 5D calculus does
not take into account certain aspects of the structure of time
translations in $\kappa$-Minkowski. We propose a change of basis
for the 5D calculus which leads to a more intuitive
description of time translations.
\end{abstract}
\maketitle

\newpage\baselineskip12pt plus .5pt minus .5pt \pagenumbering{arabic} %
\pagestyle{plain}

\section{Introduction and summary}\label{par:intro}
\noindent A rather sizeable literature has been devoted over these
past few years (see, {\it e.g.},
Refs.~\cite{dsr1dsr2, jurekmax, leedsr} and references
therein) to the possibility that the short-distance (Planck-scale)
structure of spacetime, which according to a popular
``quantum-gravity
intuition''~\cite{mead, padma, ng1994, gacmpla, garay} should be
highly nontrivial, might be such to require a new description of
spacetime symmetries. In particular, the quantum-gravity
version~\cite{gacMinkLim} of Minkowski spacetime, which should be
like classical Minkowski spacetime for soft probes but should
reveal additional structures when probed with sensitivities
approaching the Planck length $L_p$ ($L_p=\sqrt{\hbar G/c^3} \sim
10^{-35}m$), might require some deformation of the Poincar\'e
symmetries. Unfortunately, several grey areas remain in the
formalization of this intriguing hypothesis.
 A nonclassical spacetime which
several authors have considered as a candidate for the emergence of
modified spacetime symmetries is the $\kappa$-Minkowski
noncommutative spacetime~\cite{majrue, kpoinap}, with the
characteristic noncommutativity
\begin{eqnarray}
&[{x}_j,{x}_0]= i \lambda {x}_j \\
&[{x}_k,{x}_j]=0 ~,\label{kmnoncomm}
\end{eqnarray}
where ${x}_0$ is the time coordinate, ${x}_j$ are space
coordinates ($j,k \in \{1,2,3\}$), and $\lambda$ is a length scale,
usually expected to be of the order of the Planck length.

 $\kappa$-Minkowski is considered a promising candidate
 because it admits a ``natural" formulation~\cite{majrue,kowaorder,aadluna}
of rules of transformation under translations, space-rotations and
boosts, and the generators of these transformations can be
described as generators of the so-called ``$\kappa$-Poincar\'e"
Hopf algebra~\cite{lukieIW, majrue, kpoinap} rather than a standard
Lie algebra. The Hopf-algebra structures intervene primarily in
the law of action of generators on products of functions, since
the noncommutativity of the coordinates turns out to be
incompatible with the Lie-algebra requirement $T(f g)=T(f)g+fT(g)$
(if $f$ and $g$ are function of the $\kappa$-Minkowski coordinates
and $T$ is one of the generators of the transformations).

While it is easy to show that  ``$\kappa$-Poincar\'e" structures
can be naturally used to describe some mathematical features of
field theories in the  $\kappa$-Minkowski spacetime, it has been
frustratingly difficult to establish if and how these structures
affect the physics of these theories, in spite of more than a
decade of studies by several research groups. Until not long ago
one could still rightfully argue (as done, {\it e.g.}, in
Ref.~\cite{kosiNOsymm}) that the Hopf-algebra structures might
provide only a fancy mathematical formalization of a rather
trivial break down of symmetry. However, some of us recently
reported~\cite{kappanoether} a first successful Noether analysis
for some ``$\kappa$-Poincar\'e" symmetries of $\kappa$-Minkowski,
and clearly the conserved charges
derived in Ref.~\cite{kappanoether},
even though we only
managed to obtain them for the translation sector (and we did not
provide a ``measurement theory" for them), represent a significant step
forward in the direction of establishing that indeed the
$\kappa$-Poincar\'e structures encountered in the description
 of field theories in the  $\kappa$-Minkowski spacetime really  amount
 to a manifestation of a short-distance deformation of physical/observable symmetries.

The key ingredient of the Noether analysis reported in
Ref.~\cite{kappanoether} is the introduction of
translation-transformation parameters with appropriate nontrivial
commutators with the spacetime coordinates. In looking for a
suitable description of translation transformations in 4D
$\kappa$-Minkowski spacetime, Ref.~\cite{kappanoether} naturally
assumed that 4 such parameters would be needed, and indeed the
Noether analysis was successful adopting a description of the
action of translation transformations on functions of the type $f
\rightarrow f+df$, with ($\mu \in \{0,1,2,3\}$)
\begin{equation}
df= i \epsilon_{\mu}{P}^{\mu} f({x}) ~, \label{trar}
\end{equation}
where the operators ${P}^{\mu}$ are the previously known
``generators of the Majid-Ruegg basis of
$\kappa$-Poincar\'e"~\cite{majrue,kpoinap}, while for the four
``noncommutative transformation parameters" $\epsilon_{\mu}$
it turns out~\cite{kappanoether} that one must
introduce product rules with the spacetime coordinates that are
modelled on the structure of a 4D differential calculus known for
$\kappa$-Minkowski~\cite{4dDiffCalc}, amounting to the requirements
$[\epsilon_{j},{x}_0]= i \lambda \epsilon_{j}$, $[\epsilon_{j},
{x}_{k}]=0$,
 $[\epsilon_{0}, {x}_{\mu}]=0$.

More recently, in Ref.~\cite{kowafreidel}, it was argued that it
should also be possible to perform a Noether analysis of
translation transformations in 4D $\kappa$-Minkowski using a
description of translations inspired by the structure of a known
5D differential calculus for $\kappa$-Minkowski~\cite{Sitarz}.
Indeed, the possibility of
a 5D differential calculus
for a 4D spacetime, which one would not usually consider,
is rather well established in the $\kappa$-Minkowski
literature. While in classical Minkowski spacetime one has only one
natural differential calculus, there is more
then one consistent differential calculus for
$\kappa$-Minkowski, and,
besides some 4D differential calculi,
the structure of $\kappa$-Minkowski is also compatible with the
introduction of a certain 5D differential calculus~\cite{Sitarz},
the so-called ``bicovariant differential calculus".

In light of these considerations the argument put forward in
Ref.~\cite{kowafreidel} is rather reasonable.
Still for a better understanding of this
possibility of using a 5D differential calculus
for a 4D spacetime it may be useful to perform a more in-depth
analysis then the one reported in Ref.~\cite{kowafreidel}.
Moreover, since these are our first experiences with the workings
of the Noether theorem when applied to Hopf-algebra spacetime symmetries,
it is intriguing that in comparison with the previous
Ref.~\cite{kappanoether} the analysis reported in Ref.~\cite{kowafreidel}
chooses a significantly different set up for the Noether analysis.
Are these difference dictated by the structure of the 5D differential calculus?
or could one analyze the case with the 5D calculus using the same
setup that proved successful in Ref.~\cite{kappanoether}?

In particular,
Ref.~\cite{kowafreidel} relied on a proposed equivalence
between a free $\kappa$-Minkowski field theory and a free relativistically
invariant (non-local) field theory on classical Minkowski
space-time. And in setting up this commutative-theory description
a carefully taylored Weyl map was adopted.
These represent significant differences with respect to
the previous example of Noether analysis reported in Ref.~\cite{kappanoether},
which only relied on direct explicit manipulations of noncommutative fields
and did not appear to prefer any particular choice of Weyl map.

We here expose several issues that invite further scrutiny of the
 equivalence of
theories  proposed in Ref.~\cite{kowafreidel}, and in particular
we observe that this proposed equivalence associates time
translations of the commutative theory to transformations that are
not time translations in the noncommutative theory. We have
therefore chosen not to rely on attempts of reformulation in terms
of a commutative theory, but proceed as  in
Ref.~\cite{kappanoether}, by working directly in terms of the
noncommutative fields. We find that the same tools developed in
Ref.~\cite{kappanoether} for the 4D-calculus analysis are also
well suited for dealing with the 5D-calculus analysis.

Our motivation goes beyond establishing the viability of the odd
possibility of 5-parameter translation transformations in a 4D
spacetime. We are mostly hoping to contribute to the understanding
of spacetime symmetries in noncommutative spacetimes, and
particularly $\kappa$-Minkowski. The fact that there are different
(and not clearly equivalent) ways to
describe the translation transformations of a given noncommutative
spacetime is certainly very significant for the task of finding
the correct physical/operative meaning of these symmetry
transformations.
By working explicitly in terms of the fields
of interest, functions of the noncommutative $\kappa$-Minkowski
spacetime coordinates, we manage to uncover several structures that
were beyond the reach of the commutative-theory reformulation
adopted in Ref.~\cite{kowafreidel}, and could be very significant for
a full characterization of our 5-parameter transformations.
In particular we find that within the 5D-calculus setup some subtleties
must be handled when trying to establish the time independence of a noncommutative
field and at present it is not possible to formulate even a tentative proposal
of identification of the energy observable.
More material for a debate on the physical interpretation of this framework
is provided by the fact that we find 5 conserved charges from the Noether analysis
of our 5-parameter transformations.

In the next section we set up the description of our 5 parameter
transformations. Then in Section \ref{par:NoetherAnalysis} we
perform the Noether analysis, we show how time derivatives are to
be formulated in the 5D-calculus setup, and obtain 5
time-independent charges. While the analysis in Section
\ref{par:NoetherAnalysis} implicitly assumes the fields to be
real, in Section \ref{par:Dispersionrelation} we provide
generalized formulas applicable to complex fields. In Section V we
propose a change of basis for the 5D calculus which reflects our
findings on the description of time derivatives. In Section
\ref{par:Remarks} we comment on the fate of classical symmetries
within our framework. In Section \ref{par:Comparison} we compare
our findings with the ones reported in Ref.~\cite{kowafreidel2},
by the same authors of Ref.~\cite{kowafreidel}, which appeared
while we were in the final stages of preparation of this
manuscript. On several issues Ref.~\cite{kowafreidel2} adopts a
perspective which is somewhat different from the one adopted in
Ref.~\cite{kowafreidel}, but still there are very significant
differences with respect to the analysis we are reporting in this
manuscript. Section \ref{par:ClosingRemarks} offers some closing
remarks.

\section{5-parameter translations for 4D  $\kappa$-Minkowski}\label{par:DifferentialCalculus}
The argument put forward in Ref.~\cite{kowafreidel} is centered on a
well-known peculiarity of $\kappa$-Minkowski, {\it i.e.} the
availability of a ``natural"\footnote{This 5D differential calculus
is natural in the sense that it can be singled out by a
criterion~\cite{Sitarz} of ``bicovariance", which essentially
amounts to demanding that the commutators between elements of the
differential calculus and spacetime coordinates be covariant under
the action of the generators of the  $\kappa$-Poincar\'e Hopf
algebra.} 5D differential calculus $\{\hat{d}{x}_0,\hat{d}{x}_j,
\hat{d}{x}_4\}$ defined by the following commutation relations\footnote{We use
capital latin letters for indices running over $\{0,\,1,\,2,\,3,\,4\}$,
small latin letters for indices running over $\{1,\,2,\,3\}$,
and greek letters for indices running over $ \{0,\,1,\,2,\,3\}$.}
\begin{equation*}
[{x}_0, \hat{d}{x}_4] = i \lambda \hat{d}{x}_0 ~,~~~~ [{x}_0,
\hat{d}{x}_0] = i \lambda \hat{d}{x}_4 ~,~~~~ [{x}_0, \hat{d}{x}_j] = 0
\end{equation*}
\begin{equation}\label{eq:commutators}
[{x}_j, \hat{d}{x}_4] = [{x}_j, \hat{d}{x}_0] = -i \lambda
\hat{d}{x}_j ~,~~~~ [{x}_j, \hat{d}{x}_k] = i \lambda \delta_{jk}
(\hat{d}{x}_4 - \hat{d}{x}_0) ~.
\end{equation}
Assuming that the translation-transformation
parameters $\{\hat{\epsilon}_0,\hat{\epsilon}_j, \hat{\epsilon}_4\}$
are the elements of the 5D differential calculus
($\hat{\epsilon}_0\equiv \hat{d}{x}_0$, $\hat{\epsilon}_j\equiv
\hat{d}{x}_j$, $\hat{\epsilon}_4\equiv \hat{d}{x}_4$) one can
introduce infinitesimal translations of fields as maps $\Phi
\rightarrow \Phi + \hat{d} \Phi$ with
\begin{equation}\label{eq:df}
\hat{d} \Phi = i\,(\,\hat{\epsilon}^0 \hat{P}_0 + \hat{\epsilon}^j
\hat{P}_j +\hat{\epsilon}^4 \hat{P}_4\,)\, \Phi
\end{equation}
where the operators $\hat{P}_0, \hat{P}_j, \hat{P}_4$ are simply
related to the operators ${P}_0, {P}_j$ used by some of us in the
4-parameter description of translations reported in
Ref.~\cite{kappanoether}:
\begin{eqnarray}
&\hat{P}_0 = \frac{1}{\lambda} ( \sinh {\lambda P_0} +
\frac{\lambda^2}{2} \vec{P}^2 e^{ \lambda P_0} )\nonumber\\
&\hat{P}_i = P_i e^{ \lambda P_0}\nonumber\\
&\hat{P}_4 = \frac{1}{\lambda} ( \cosh {\lambda P_0} -1 -
\frac{\lambda^2}{2} \vec{P}^2 e^{ \lambda P_0} ) \label{eq:generatori}
\end{eqnarray}

The ${P}_0, {P}_j$ used in Ref.~\cite{kappanoether}
are the translation generators of the Majid-Ruegg basis
of the $\kappa$-Poincar\'e Hopf algebra~\cite{majrue,kpoinap},
while the $\hat{P}_0, \hat{P}_j$ of (\ref{eq:df})
clearly provide a different basis of generators for $\kappa$-Poincar\'e.
It is easy to verify~\cite{majrue,aadluna} that
the fifth operator, $\hat{P}_4$, is a casimir of
the $\kappa$-Poincar\'e Hopf-algebra~\cite{majrue,kpoinap}.

Since one can write any $\kappa$-Minkowski field $\Phi({x})$
in the form~\cite{wessDefinizio,aadluna}
\begin{eqnarray}
\Phi({x})=\int\,d^4p\,\tilde{\Phi}(p)\,e^{i\vec{p}\cdot
\vec{x}}\,e^{-i{p}_0{x}_0} ~, \label{fourier}
\end{eqnarray}
in order to provide an explicit description of the action of the Majid-Ruegg
generators it is sufficient to specify that
\begin{eqnarray}
P_{\mu}\triangleright \left( e^{i\vec{k}\cdot \vec{x}}\,e^{-i{k}_0{x}_0} \right) \,=
k_\mu\,e^{i\vec{k}\cdot \vec{x}}\,e^{-i{k}_0{x}_0}\,  \nonumber
\end{eqnarray}
and accordingly for the $\hat{P}_A$ operators one finds
\begin{eqnarray}
\hat{P}_0\triangleright \left( e^{i\vec{k}\cdot \vec{x}}\,
e^{-i\vec{k}_0{x}_0} \right) \, &=& \frac{1}{\lambda} ( \sinh {\lambda k_0} +
\frac{\lambda^2}{2} \vec{k}^2 e^{ \lambda k_0} )\,
e^{i\vec{k}\cdot \vec{x}}\,e^{-i{k}_0{x}_0}\,, \nonumber\\
\hat{P}_i \triangleright \left( e^{i\vec{k}\cdot \vec{x}}\,
e^{-i{k}_0\,{x}_0} \right) \, &=& k_i \,e^{ \lambda k_0}\, \,
e^{i\vec{k}\cdot \vec{x}}\,e^{-i{k}_0{x}_0}\, ,\nonumber\\
\hat{P}_4\triangleright \left( e^{i\vec{k}\cdot \vec{x}}\,
e^{-i{k}_0{x}_0} \right) \, &=& \frac{1}{\lambda} ( \cosh {\lambda k_0}-1
- \frac{\lambda^2}{2} \vec{k}^2 e^{ \lambda k_0} )\, e^{i\vec{k}\cdot
\vec{x}}\,e^{-i{k}_0{x}_0}\, . \label{actionsJOC}
\end{eqnarray}

In light of the form of these rules of action we should explicitly warn
the reader of the fact that our notation is potentially misleading: usually
a $0$ index on a $P$ generator identifies a time-translation generator,
but clearly our $\hat{P}_0$ does not generate time translations (in particular,
it does not vanish on time-independent functions).
We shall take this into account (actually the formalism itself will
direct us toward the correct way to handle this issue) and comment again on
this point when appropriate.

In preparation for our Noether analysis it is useful to observe that
(using the rules (\ref{actionsJOC}) of action of the $\hat{P}_A$
operators and the rules of commutation between transformation
parameters and $\kappa$-Minkowski coordinates) one easily
verifies~\cite{aadluna, kappadirac, dandi, anto} that the
differential $\hat{d} \Phi$ defined in (\ref{eq:df}) satisfies
Leibniz rule
\begin{equation}\label{liebn}
\hat{d} (\Phi \Psi) = \Phi (\hat{d}  \Psi) + (\hat{d} \Phi ) \Psi\,.
\end{equation}

For the Noether analysis it is also useful to notice that from the
rules of commutation between transformation parameters (elements of
the 5D differential calculus) and $\kappa$-Minkowski coordinates one
obtains the following rules of commutation between
transformation parameters
 and time-to-the-right-ordered exponentials:
\begin{eqnarray}\label{eq:planewave}
&e^{i\vec{k} \cdot \vec{x}}e^{- i k_0 {x}_0}\,  \hat{\epsilon}_0 = \left( (
 \lambda \hat{P}_0+  e^{ -\lambda P_0} )  \hat{\epsilon}_0 + \lambda
 \hat{P_i} \hat{\epsilon}_i
+(\lambda \hat{P}_4+1-  e^{ -\lambda P_0} )
 \hat{\epsilon}_4 \right) e^{i\vec{k} \cdot \vec{x}}e^{- i k_0 {x}_0}\nonumber\\
&e^{i\vec{k} \cdot \vec{x}}e^{- i k_0 {x}_0}\, \hat{\epsilon}_i = \left(
 \lambda e^{ -\lambda P_0} \hat{P}_i \hat{\epsilon}_0 + \hat{\epsilon}_i - \lambda e^{ -\lambda P_0} \hat{P}_i
  \hat{\epsilon}_4 \right) e^{i\vec{k} \cdot \vec{x}}e^{- i k_0 {x}_0}\nonumber\\
&e^{i\vec{k} \cdot \vec{x}}e^{-i k_0 {x}_0}\,  \hat{\epsilon}_4 = \left(
\lambda \hat{P}_0  \hat{\epsilon}_0 + \lambda
 \hat{P}_i \hat{\epsilon}_i + (\lambda \hat{P}_4+1)
 \hat{\epsilon}_4 \right) e^{i\vec{k} \cdot \vec{x}}e^{-i k_0 {x}_0}~.
\end{eqnarray}
\noindent
And it is useful to write down the ``coproducts" of the
operators $\hat{P}_A$,\footnote{In writing these coproducts we found convenient, in order
to keep equations short, to let intervene the operator $P_0$. In order to write
these coproducts exclusively in terms of the operators $\hat{P}_A$ one can of course use the
fact that from (\ref{eq:generatori}) it follows that $\exp(\lambda P_0)=\lambda (\hat{P}_0+\hat{P}_4)+1$.}
\begin{equation}\label{eq:coprodotti1}
\Delta (\hat{P}_0 ) = \hat{P}_0 \otimes e^{\lambda P_0} +
e^{-\lambda P_0} \otimes \hat{P}_0 + \lambda e^{-\lambda
P_0}\hat{P}_i \otimes \hat{P}_i\,,
\end{equation}
\begin{equation}\label{eq:coprodotti2}
\Delta (\hat{P}_i ) = \hat{P}_i \otimes e^{\lambda P_0} + 1 \otimes
\hat{P}_i\,,
\end{equation}
\begin{equation}\label{eq:coprodotti3}
\Delta (\hat{P}_4 ) = \hat{P}_4 \otimes e^{\lambda P_0} -
e^{-\lambda P_0} \otimes \hat{P}_0 - \lambda e^{-\lambda
P_0}\hat{P}_i \otimes \hat{P}_i + 1\otimes (\hat{P}_0+\hat{P}_4)\,,
\end{equation}
which describe in standard Hopf-algebra notation the action of the operators $\hat{P}_A$
on products of functions. For example, Eq.~(\ref{eq:coprodotti2})
reflects the fact that from (\ref{actionsJOC})
it follows that
\begin{eqnarray*}
&\hat{P}_i\,\left( e^{i\vec{k} \cdot \vec{x}}e^{-i  k_0 {x}_0}\cdot
e^{i\vec{p} \cdot \vec{x}}e^{- i p_0 {x}_0}\right)=\hat{P}_i\,\left(e^{i(\vec{k}+
e^{-\lambda k_0}\vec{p}) \cdot \vec{x}}e^{- i (k_0+p_0) {x}_0}\right)= \\
&=({k_i}+e^{-\lambda k_0}{p_i})\, e^{\lambda (k_0+p_0)}\,\left(e^{i(\vec{k}+
e^{-\lambda k_0}\vec{p}) \cdot \vec{x}}e^{- i(k_0+p_0) {x}_0}\right)=      \\
&=({k_i}e^{\lambda(k_0+ p_0)}+e^{\lambda p_0}{p_i})\,\left( e^{i(\vec{k}+
e^{-\lambda k_0}\vec{p}) \cdot \vec{x}}e^{-i (k_0+p_0) {x}_0}\right)=\\
&=\hat{P}_i\left( e^{i\vec{k} \cdot \vec{x}}e^{- ik_0 {x}_0} \right)\cdot
e^{\lambda P_0}\left(  e^{i\vec{p} \cdot \vec{x}}e^{- ip_0 {x}_0}\right)+
\left( e^{i\vec{k} \cdot \vec{x}}e^{- i k_0 {x}_0} \right)\cdot \hat{P}_i \left(  e^{i\vec{p} \cdot \vec{x}}
e^{- i p_0 {x}_0}\right)\,.
\end{eqnarray*}

\section{Noether analysis}\label{par:NoetherAnalysis}
We are now ready to test our 5-parameter transformations within a
Noether analysis for classical fields in the noncommutative
$\kappa$-Minkowski spacetime. As mentioned in the Introduction it is
interesting for us to verify whether this analysis can be performed
following exactly the techniques already developed in
Ref.~\cite{kappanoether} or it is necessary to adapt the procedure
to some peculiarities of the 5D differential calculus. We shall
therefore proceed exactly as in Ref.~\cite{kappanoether} and deduce
from the success of this Noether analysis that it is not necessary
to adapt the Noether analysis to the type of differential calculus
in use. As basis for our illustrative example of Noether analysis we
consider one of the most studied equations of motion in the
$\kappa$-Minkowski literature~\cite{kpoinap, aadluna} which concerns
a scalar field $\Phi({x})$ governed by the Klein-Gordon-like
equation of motion
\begin{equation}\label{eq:5DMotionEquation}
C_\lambda(P_\mu)\,\Phi \equiv \left[\left(\frac{2}{\lambda} \sinh
{\frac{\lambda}{2} P_0}\right)^2-e^{\lambda
P_0}\vec{P}^2\right]\Phi=m^2\Phi ~,
\end{equation}
whose most general solution can be written in the form\footnote{We consider
the function $C_\lambda$,
defined in (\ref{eq:5DMotionEquation}), sometimes on operators, as in the case
of $C_\lambda(P_\mu)$ which is an operator, and sometimes on Fourier parameters,
as in the case of $C_\lambda(k_\mu)$ which is just a number obtained from the
Fourier parameters.}
\begin{equation} \label{eq:Solutionm=0}
\Phi({x})=\int d^4k\, \tilde{\Phi}(k_0,\vec{k})\,e^{i\vec{k}
\cdot \vec{{x}}}e^{-i
k_0{x}_0}\,\delta(C_\lambda(k_\mu)-m^2).
\end{equation}

Our Noether analysis relies on the fact that this equation of motion
can be derived~\cite{kappanoether} from the following action
\begin{eqnarray}\label{eq:Action}
S[\Phi]&=&\int d^4x \mathcal{L}[\Phi({x})]\nonumber\\
\mathcal{L}[\Phi({x})]&=&\frac{1}{2} \left(\Phi({x})
\,C_\lambda\,\Phi({x}) - m^2\Phi({x})\Phi({x})\right)
\end{eqnarray}
\noindent
The operator $C_\lambda(P_\mu)$ defined in (\ref{eq:5DMotionEquation})
commutes with all the operators introduced in the previous section,
since it is a Casimir of the $\kappa$-Poincar\'{e} Hopf algebra.
We find sometimes useful to also write\footnote{As customary,
we adopt the Einstein summation rule for greek and latin indices.}
it as $C_\lambda=\tilde{P}_\mu\tilde{P}^\mu$
in terms of the operators
\begin{equation} \label{eq:Ptilde}
\tilde{P}_0 = \frac{2}{\lambda} \sinh {\frac{\lambda}{2} P_0}\,, \qquad
\tilde{P}_i = P_i e^{\frac{\lambda}{2} P_0}\, ,
\end{equation}
which turn out to allow to write more compactly some of the equations.

We are of course interested in analyzing the variation of the
Lagrangian density under our 5-parameter transformation. This can be
done easily using Eqs.~(\ref{eq:df}) and (\ref{eq:planewave}),
noting that
\begin{equation}
\tilde{P_\alpha}[f({x})g({x})]=[\tilde{P_\alpha}f({x})][e^{\frac{\lambda}{2}P_0}g({x})]+
[e^{-\frac{\lambda}{2}P_0}f({x})][\tilde{P_\alpha}g({x})]\,,
\end{equation}\\
\noindent and using the fact that, by definition of a scalar field
(and assuming $\delta\Phi({x}')\equiv
\Phi'({x}')-\Phi({x}')\simeq
\Phi'({x})-\Phi({x})\equiv\delta\Phi({x})$)
\begin{equation}
0=\Phi'({x}')-\Phi({x})=[\Phi'({x}')-\Phi({x}')]-[\Phi({x}')-\Phi({x})]~,
\end{equation}
{\it i.e.} $\delta\Phi=-\hat{d}\Phi =-i\left( \,\hat{\epsilon}^0
\hat{P}_0 + \hat{\epsilon}^j \hat{P}_j +\hat{\epsilon}^4 \hat{P}_4\,
\right)\, \Phi$.

The result (whose detailed derivation will be reported
elsewhere~\cite{dandi,anto}) is
\begin{equation*}
0=\delta\mathcal{L}=\frac{1}{2}  \left( \delta\Phi\, C_\lambda \Phi
+\Phi\, C_\lambda\, \delta\Phi -
m^2\delta\Phi\,\Phi-m^2\Phi\,\delta\Phi\right)=
\end{equation*}
\begin{equation}
= -\frac{1}{2} \left\{e^{\frac{\lambda
P_0}{2}}\tilde{P}^0\left[\left(\frac{2}{\lambda}+\lambda
m^2-\frac{e^{\lambda P_0}}{\lambda}\right)\Phi\,\delta\Phi-\Phi
\frac{e^{-\lambda
P_0}}{\lambda}\delta\Phi\right]+\hat{P}^i\left[\Phi e^{-\lambda
P_0}\hat{P}_i\delta\Phi-\hat{P}_i\Phi\,\delta\Phi\right]\right\}\, , \label{variazio}
\end{equation}
where we already specialized to the case of fields such
that $\tilde{P}^\mu \tilde{P}_\mu\Phi=m^2\Phi$, since of course we perform the
Noether analysis on fields that are solutions of the equation of motion.

In (\ref{variazio}) the transformation parameters $\hat{\epsilon}_A$
appear implicitly through $\delta\Phi$.
It is convenient to use the formulas (\ref{eq:planewave}) to carry
all the $\hat{\epsilon}_A$ to the left side of the monomials
composing the expression of $\delta\mathcal{L}$. This allows to rewrite Eq.~(\ref{variazio})
in the form~\cite{dandi, anto}
\begin{equation}\label{protodiv}
\hat{\epsilon}^A\left(\,e^{\frac{\lambda P_0}{2}}\tilde{P}^0\,J_{0A}+\hat{P}^iJ_{iA}\right)=0\,,
\end{equation}
where
\begin{eqnarray}
J_{00}&=&\frac{1}{2}\bigg\{\left(\frac{2}{\lambda}+\lambda
m^2-\frac{e^{\lambda P_0}}{\lambda}\right)\left[(\lambda\hat{P}_0+
e^{ -\lambda P_0} )\Phi\hat{P}_0\Phi + \lambda
 P_i\Phi\hat{P_i}\Phi
+\lambda \hat{P}_0\Phi\hat{P}_4\Phi\right]+\nonumber\\
\!\!&-&\!\!(\lambda\hat{P}_0+ e^{ -\lambda P_0} )\Phi
\frac{e^{-\lambda P_0}}{\lambda}\hat{P}_0\Phi - \lambda
 P_i\Phi\frac{e^{-\lambda P_0}}{\lambda}\hat{P_i}\Phi-
\lambda \hat{P}_0\Phi\frac{e^{-\lambda P_0}}{\lambda}\hat{P}_4\Phi\bigg\}\,, \nonumber \\
\nonumber \\
J_{0i}&=&\frac{1}{2}\bigg\{\left(\frac{2}{\lambda}+\lambda
m^2-\frac{e^{\lambda
P_0}}{\lambda}\right)\bigg[\lambda\hat{P}_i\Phi\hat{P}_0\Phi+\Phi\hat{P}_i\Phi+\lambda\hat{P}_i\Phi\hat{P}_4\Phi\bigg]+\nonumber\\
&-&\,\lambda\hat{P}_i\frac{e^{-\lambda
P_0}}{\lambda}\Phi\hat{P}_0\Phi-\Phi\hat{P}_i\frac{e^{-\lambda
P_0}}{\lambda}\Phi-\lambda\hat{P}_i\Phi\frac{e^{-\lambda
P_0}}{\lambda}\hat{P}_4\Phi\bigg\}\,, \nonumber\\
\nonumber \\
J_{04}&=&\frac{1}{2}\bigg\{\bigg(\frac{2}{\lambda}+\lambda
m^2-\frac{e^{\lambda P_0}}{\lambda}\bigg)\bigg[(\lambda\hat{P}_4+1-
e^{ -\lambda P_0} )\Phi\hat{P}_0\Phi - \lambda
 P_i\Phi\hat{P_i}\Phi
+(\lambda\hat{P}_4+1)\Phi\hat{P}_4\Phi\bigg]+\nonumber\\
&-&(\lambda\hat{P}_4+1-
e^{-\lambda P_0} )\Phi \frac{e^{-\lambda P_0}}{\lambda}\hat{P}_0\Phi +\lambda
 P_i\Phi\frac{e^{-\lambda P_0}}{\lambda}\hat{P_i}\Phi-
 (\lambda\hat{P}_4+1)\Phi\frac{e^{-\lambda
P_0}}{\lambda}\hat{P}_4\Phi\bigg\}\,. \label{eq:J}
\end{eqnarray}
First let us notice that the Noether analysis leads automatically to a structure
that does not require terms of the type $\hat{P}^4J_{4A}$, which is a reassuring
feature considering the peculiar nature of the operator $\hat{P}^4$ needed for
our 5-parameter transformations.
It is even more
noteworthy that the Noether analysis leads automatically to a structure
of the form $\exp({\lambda P_0}/{2})\tilde{P}^0\,J_{0A}+\hat{P}^iJ_{iA}$.
The observations we reported in Section II concerning the operator $\hat{P}^0$
imply that it would have been puzzling if the Noether analysis had led to
a structure of the type $\hat{P}^\mu J_{\mu A}$, since $\hat{P}^0$
does not vanish on time-independent fields.
The Eq.~(\ref{protodiv}) produced by the Noether analysis is at least plausible,
since, in light of the definitions
given in Section II, $\exp({\lambda P_0}/{2})\tilde{P}_0$ does
vanish on time-independent fields.
This motivates us to adopt the suggestive notation
\begin{equation}
{\mathcal{D}}_0\equiv \exp({\lambda P_0}/{2})\tilde{P}_0 ~.
\label{partial0}
\end{equation}
But the evidence of robustness of Eq.~(\ref{protodiv}) goes even
beyond ${\mathcal{D}}_0$: we find that the role played by the
structure ${\mathcal{D}}^0 J_{0A}+\hat{P}^iJ_{iA}$ in our
Noether analysis is completely analogous to the role of the
4-divergence of the currents in the Noether analysis of ordinary
theories in classical Minkowski spacetime. In order to provide
support for this statement we describe spatial integration in
$\kappa$-Minkowski as codified in the formula
\begin{equation}
\int d^3x e^{i\vec{p}\cdot \vec{x}}e^{-ip_0\,{x}_0}
=\delta(\vec{p})\,e^{-ip_0\,{x}_0}\, ,  \label{integratio}
\end{equation}
so that for a $\kappa$-Minkowski field
$\Psi({x})=\int d^4p\, \tilde{\Psi}(p_0,\vec{p})\, \exp(i\vec{p}\cdot \vec{x})\, \exp(-ip_0\,{x}_0)$
one obtains
\begin{equation}
\int d^3x \Psi({x}_0,\vec{x})= \int dp_0 \tilde{\Psi}(p_0,\vec{0})\, e^{-i p_0\,{x}_0} \,.
\end{equation}
With these rules of integration one easily obtains
from (\ref{protodiv}) (which must be valid for any arbitrary choice
of the parameters $\hat{\epsilon}^A$) the following chain of relations~\cite{dandi, anto}:
\begin{equation}
{\mathcal{D}}_0  \int d^3x \,J_{0A} = \int d^3x
\,{\mathcal{D}}_0\,J_{0A}=- \int d^3x \hat{P}^iJ_{iA} =0 ~,
\end{equation}
where on the right-hand side we used (\ref{integratio}) and the rule of action
of the operators $\hat{P}^i$ introduced in Section II.

This argument guarantees that the charges
\begin{equation}
\hat{Q}_A \equiv \int d^3x \,J_{0A}
\label{chargedefi}
\end{equation}
are indeed time independent. But of course one does not need to rely
on this argument, since the time independence of the charges can be
verified directly. This explicit verification can be done rather
straightforwardly using the techniques of Ref.~\cite{kappanoether}.
By direct evaluation of the charges one obtains a result (whose
derivation will be reported in detail elsewhere~\cite{dandi, anto})
which is indeed explicitly time independent, and can be most
conveniently expressed in terms of the Fourier transform
$\tilde{\Phi}(k)$ of a field $\Phi({x})$ solution of the
equation of motion:
\begin{equation}
\left(\begin{array}{c}
\hat{Q}_0\\
\hat{Q}_i\\
\hat{Q}_4\\
\end{array}{}\right)
=-\frac{1}{2}\int d^4k \,\tilde{\Phi}(k)
\tilde{\Phi}(\dot{-}k)\,e^{3\lambda
k_0}  \left(\begin{array}{c}
\hat{k}_0\\
\hat{k}_i\\
\hat{k}_4\\
\end{array}{}\right) \frac{(-2\tilde{k_0}e^{\frac{\lambda}{2}k_0}+\lambda
m^2)}{|-2\tilde{k_0}e^{\frac{\lambda}{2}k_0}+\lambda
m^2|}\,\delta(C_{\lambda}(k)-m^2)\,,\label{eq:Qmu}
\end{equation}
\\
where, for compactness, we introduced the notations $k \equiv (k_0,\vec{k})$,
 $\dot{-}k \equiv (-k_0, -\vec{k}e^{\lambda k_0})$, and
 (analogously to the notation $\tilde{P}_{\mu}$ previously introduced for
 frequently occurring combinations of the $P_{\mu}$ generators)
 $\tilde{k}|_{k_0,\vec{k}}\equiv \{\tilde{k}_0,\vec{\tilde{k}}\}|_{k_0,\vec{k}}\equiv
\big\{\frac{2}{\lambda}\sinh(\frac{\lambda}{2} k_0),\vec{k} \exp(\frac{\lambda}{2} k_0)\big\}$, as well as $\{\hat{k}_0,\,\hat{k}_i,\,\hat{k}_4\}|_{k_0,\vec{k}}\equiv\big\{\frac{1}{\lambda} ( \sinh {\lambda k_0} +
\frac{\lambda^2}{2} \vec{k}^2 e^{ \lambda k_0} )\,,k_i e^{ \lambda k_0},\, \frac{1}{\lambda} ( \cosh {\lambda k_0} -1 -
\frac{\lambda^2}{2} \vec{k}^2 e^{ \lambda k_0} )\big\}$.

Besides being time independent, for real fields (and complex fields, but this we discuss in the
next section) our charges are also automatically real, as one
can verify using the fact that a
scalar field $\Phi({x})$ solution of our equation of
motion $C_{\lambda}(P)\,\Phi=m^2\Phi$ on $\kappa$-Minkowski,
\begin{equation*}
\Phi({x})=\int
d^4k\,\tilde{\Phi}(k)\,\delta(C_{\lambda}(k)-m^2)e^{i\vec{k}\cdot\vec{{x}}}e^{-ik_0\, {x}_0}
~,
\end{equation*}
will be real if
\begin{equation}\label{eq:reality}
\tilde{\Phi}(k_0,\vec{k})=\left(\tilde{\Phi}(-k_0,-\vec{k}e^{\lambda
k_0})\right)^*e^{3\lambda k_0}\,,
\end{equation}
and this allows us to rewrite the charges (\ref{eq:Qmu}) as explicitly real\footnote{We shall later contemplate the possibility of on-shell fields with complex $k_0$ Fourier parameter. Note, however, that $\hat{k}_0,\,\hat{k}_i,\,\hat{k}_4$ and $\tilde{k}_0e^{\frac{\lambda}{2}k_0}$ are real even when $k_0$ has an imaginary part. } quantities:
\begin{equation}\label{qreal}
\left(\begin{array}{c}
\hat{Q}_0\\
\hat{Q}_i\\
\hat{Q}_4\\
\end{array}{}\right)
=-\frac{1}{2}\int  d^4k
\left|\tilde{\Phi}(k)\right|^2    \left(\begin{array}{c}
\hat{k}_0\\
\hat{k}_i\\
\hat{k}_4\\
\end{array}{}\right) \, \frac{(-2\tilde{k_0}e^{\frac{\lambda}{2}k_0}+\lambda
m^2)}{|-2\tilde{k_0}e^{\frac{\lambda}{2}k_0}+\lambda
m^2|}\,\delta(C_{\lambda}(k)-m^2)\,.
\end{equation}

\section{The case of complex scalar fields} \label{par:Dispersionrelation}
In the previous section we considered real scalar fields, but
actually the steps of the analysis only require little or no
adaptation for the case of complex fields. Essentially it reduces to
the fact that in appropriate places one must consider the complex
conjugate  $\Phi^*({x})$ of the field $\Phi({x})$. For a complex
field (which really amounts to 2 real fields) most authors would
also remove the factor $1/2$ in front of the action, besides
introducing the characteristic structure ``$\Phi^* \dots \Phi$":
\begin{eqnarray}
S[\Phi]&=&\int d^4x \mathcal{L}[\Phi({x})]\nonumber\\
\mathcal{L}[\Phi({x})]&=& \left(\Phi^*({x})
\,C_\lambda\,\Phi({x}) -
m^2\Phi^*({x})\Phi({x})\right)\,.
\end{eqnarray}

Proceeding exactly in the same way as in the previous section one
then easily arrives once again to the equation
$\hat{\epsilon}^A\left(\,e^{\frac{\lambda
P_0}{2}}\tilde{P}^0\,J_{0A}+\hat{P}^iJ_{iA}\right)=0$, with $J$'s of
the same form as in the previous section but involving
$\Phi^*({x})$ in appropriate places. In particular, one finds
\begin{eqnarray}
J_{00}&=&\bigg\{\left(\frac{2}{\lambda}+\lambda
m^2-\frac{e^{\lambda P_0}}{\lambda}\right)\left[(\lambda\hat{P}_0+
e^{ -\lambda P_0} )\Phi^*\hat{P}_0\Phi + \lambda
 P_i\Phi^*\hat{P_i}\Phi
+\lambda \hat{P}_0\Phi^*\hat{P}_4\Phi\right]+\nonumber\\
\!\!&-&\!\!(\lambda\hat{P}_0+ e^{ -\lambda P_0} )\Phi^*
\frac{e^{-\lambda P_0}}{\lambda}\hat{P}_0\Phi - \lambda
 P_i\Phi^*\frac{e^{-\lambda P_0}}{\lambda}\hat{P_i}\Phi-
\lambda \hat{P}_0\Phi^*\frac{e^{-\lambda P_0}}{\lambda}\hat{P}_4\Phi\bigg\}\,, \nonumber \\
\nonumber \\
J_{0i}&=&\bigg\{\left(\frac{2}{\lambda}+\lambda
m^2-\frac{e^{\lambda
P_0}}{\lambda}\right)\bigg[\lambda\hat{P}_i\Phi^*\hat{P}_0\Phi+\Phi^*\hat{P}_i\Phi+\lambda\hat{P}_i\Phi^*\hat{P}_4\Phi\bigg]+\nonumber\\
&-&\,\lambda\hat{P}_i\frac{e^{-\lambda
P_0}}{\lambda}\Phi^*\hat{P}_0\Phi-\Phi^*\hat{P}_i\frac{e^{-\lambda
P_0}}{\lambda}\Phi-\lambda\hat{P}_i\Phi^*\frac{e^{-\lambda
P_0}}{\lambda}\hat{P}_4\Phi\bigg\}\,, \nonumber\\
\nonumber \\
J_{04}&=&\bigg\{\bigg(\frac{2}{\lambda}+\lambda
m^2-\frac{e^{\lambda P_0}}{\lambda}\bigg)\bigg[(\lambda\hat{P}_4+1-
e^{ -\lambda P_0} )\Phi^*\hat{P}_0\Phi - \lambda
 P_i\Phi^*\hat{P_i}\Phi+(\lambda\hat{P}_4+1)\Phi^*\hat{P}_4\Phi\bigg]+
\nonumber\\
&-&(\lambda\hat{P}_4+1-
e^{-\lambda P_0} )\Phi^* \frac{e^{-\lambda P_0}}{\lambda}\hat{P}_0\Phi +\lambda
 P_i\Phi^*\frac{e^{-\lambda P_0}}{\lambda}\hat{P_i}\Phi-
 (\lambda\hat{P}_4+1)\Phi^*\frac{e^{-\lambda
P_0}}{\lambda}\hat{P}_4\Phi\bigg\}\,. \label{eq:Jcomplex}
\end{eqnarray}\\
And from these one obtains the time-independent charges, which are also conveniently
written in terms of Fourier transforms:
\begin{equation}\label{eq:QmuComplex}
\left(\begin{array}{c}
\hat{Q}_0\\
\hat{Q}_i\\
\hat{Q}_4\\
\end{array}{}\right)
=-\int d^4k \,\tilde{\Phi}(k)
\tilde{\Phi}^*(\dot{-}k)\,e^{3\lambda
k_0}  \left(\begin{array}{c}
\hat{k}_0\\
\hat{k}_i\\
\hat{k}_4\\
\end{array}{}\right) \frac{(-2\tilde{k_0}e^{\frac{\lambda}{2}k_0}+\lambda
m^2)}{|-2\tilde{k_0}e^{\frac{\lambda}{2}k_0}+\lambda
m^2|}\,\delta(C_{\lambda}(k)-m^2)\,.
\end{equation}

Also for complex fields the charges are real, as one easily verifies
using the relation~\cite{dandi,anto} between $\tilde{\Phi}$ and $\tilde{\Phi}^*$,
\begin{equation}\label{eq:Complexity}
\tilde{\Phi}(k_0,\vec{k})=\left(\tilde{\Phi}^*(-k_0,-\vec{k}e^{\lambda
k_0})\right)^*e^{3\lambda k_0}\,,
\end{equation}
which allows to rewrite the charges in the following way:
\begin{equation}\label{qcompl}
\left(\begin{array}{c}
\hat{Q}_0\\
\hat{Q}_i\\
\hat{Q}_4\\
\end{array}{}\right)
=-\int  d^4k \left|\tilde{\Phi}(k)\right|^2\, \left(\begin{array}{c}
\hat{k}_0\\
\hat{k}_i\\
\hat{k}_4\\
\end{array}{}\right) \, \frac{(-2\tilde{k_0}e^{\frac{\lambda}{2}k_0}+\lambda
m^2)}{|-2\tilde{k_0}e^{\frac{\lambda}{2}k_0}+\lambda
m^2|}\,\delta(C_{\lambda}(k)-m^2)
\end{equation}

\section{A different choice of basis for the 5D calculus}\label{par:DifferentBasis}
In this section we propose a change of basis for the 5D
differential calculus that reflects our findings concerning the
operator $\hat{P}_0$, which clearly could not be used to generate
time translations, and our operator ${\mathcal{D}}_0=\hat{P}_0+\hat{P}_4$
which instead appears to be a good candidate as generator of time translations.
In order to describe translation transformations explicitly in terms of ${\mathcal{D}}_0$ (rather than separately $\hat{P}_0$ and $\hat{P}_4$)
we propose\footnote{Our core proposal here is that an operator proportional
to ${\mathcal{D}}_0$ (or even just monotonic function of ${\mathcal{D}}_0$)
should intervene explicitly in the ``$df$ rule".
This can be done in many ways, but among these possibilities
we chose to illustrate our idea in the case where the new basis is obtained
from the $\hat{d}x_A$ basis
by action of a rotation matrix of determinant $1$.
And it must be stressed that most of the welcome features of our
new basis could be already achieved by essentially simply noticing
that $\hat{\epsilon}^A\,{\mathcal{D}}^0\,J_{0A}=
{\mathcal{D}}^0\big(\hat{\epsilon}^0(J_{00}
+J_{04})+\hat{\epsilon}^iJ_{0i}+(\hat{\epsilon}^4-\hat{\epsilon}^0)J_{04}\big)$.}
the following change of basis $\hat{d}x_A \rightarrow \bar{d}x_A$:
\begin{equation}\label{eq:dxChange}
\bar{d}x_0=(\hat{d}x_0+
\hat{d}x_4)/\sqrt{2},~~~~~~~~~\bar{d}x_i=\hat{d}x_i,
~~~~~~~~~\bar{d}x_4=(\hat{d}x_0-\hat{d}x_4)/\sqrt{2} \,.
\end{equation}
The rules of commutation between the $\bar{d}x_A$ and the
$\kappa$-Minkowski coordinates, which one easily obtains from the
corresponding rules of commutation between the $\hat{d}x_A$ and
the $\kappa$-Minkowski coordinates, take a rather simple
form:\footnote{Actually Sitarz, in Ref.~\cite{Sitarz}, already
noticed that this change of basis $\hat{d}x_A \rightarrow
\bar{d}x_A$ leads to simple commutation relations, but he had not
realized that it would also allow a more intuitive
characterization of time translations.}
\begin{equation*}
[{x}_0, \bar{d}{x}_0] = i \lambda \bar{d}{x}_0, ~~~~~~~~~   [{x}_0,
\bar{d}{x}_4] =- i \lambda \bar{d}{x}_4, ~~~~~~~~~   [{x}_0, \bar{d}{x}_j] = 0,
\end{equation*}
\begin{equation}
[{x}_j, \bar{d}{x}_4] =0,~~~~~~~~~   [{x}_j, \bar{d}{x}_0] = -\sqrt{2}\,i \lambda
\bar{d}{x}_j, ~~~~~~~~~  [{x}_j, \bar{d}{x}_k] = -\sqrt{2}\,i \lambda \delta_{jk}
\bar{d}{x}_4\, .
\end{equation}
Based on the form of this new basis for the 5D calculus one can perform
a corresponding rotation of transformation parameters,
\begin{equation}\label{eq:BasisChange}
\bar{\epsilon}_0=(\hat{\epsilon}_0+\hat{\epsilon}_4)/\sqrt{2},
~~~~~~~~~\bar{\epsilon}_i\equiv\hat{\epsilon}_i,
~~~~~~~~~\bar{\epsilon}_4=(\hat{\epsilon}_0-\hat{\epsilon}_4)/\sqrt{2},
\end{equation}
and introduce the operators
\begin{equation}
\bar{\mathcal{D}}_0 =(\hat{P}_0+\hat{P}_4)/\sqrt{2}=
\mathcal{D}_0/\sqrt{2},~~
~~~~~~~\bar{\mathcal{D}}_i\equiv\hat{P}_i,~~~~~~~~~\bar{\mathcal{D}}_4=(\hat{P}_0-\hat{P}_4)/\sqrt{2}\,,
\end{equation}
so that it is then possible to rewrite the $\hat{d}f$ we used in the previous sections
in the following way
\begin{equation}\label{eq:df2}
\hat{d}f \,=\,i\,(\bar{\epsilon}_0\bar{\mathcal{D}}_0
+\bar{\epsilon}_i\bar{\mathcal{D}}_i+\bar{\epsilon}_4\bar{\mathcal{D}}_4)f
\equiv \bar{d}f \, .
\end{equation}

It is clear from the structure of this redefinitions that by this way
to rewrite $\hat{d}f$, while allowing to introduce a parameter ($\bar{\epsilon}_0$)
that can be meaningfully described as a time-translation parameter,
one does not affect in any ``armful way" the progress of the Noether analysis.
Indeed following the same steps we described in Section \ref{par:NoetherAnalysis}, one
easily arrives at the following expression for the Lagrangian-density variation
\begin{equation}
\delta\mathcal{L}=i\,\bar{\epsilon}^A\left(\bar{\mathcal{D}}^0
\bar{J}_{0A}+ \bar{\mathcal{D}}^i \bar{J}_{iA}\right) \,,
\end{equation}
\noindent
where the quantities $\bar{J}_{\mu A}$ can be written in terms of
the $J_{\mu A}$ (for which we gave the explicit functional dependence on the field
in Section \ref{par:NoetherAnalysis}) in the following simple way:
\begin{equation}
\bar{J}_{00}\equiv J_{00}+J_{0 4},~~~~~~\bar{J}_{0i}\equiv
\sqrt{2}\, J_{0 i}, ~~~~~~\bar{J}_{04}\equiv J_{00}-J_{04}\,,\nonumber
\end{equation}
\begin{equation}
\bar{J}_{i0}\equiv (J_{i0}+J_{i4})/\sqrt{2},~~~~~~\bar{J}_{ik}\equiv
J_{ik}, ~~~~~~\bar{J}_{i 4}\equiv (J_{i0}-J_{i4})/\sqrt{2}\,.
\end{equation}
And then (proceeding again in complete analogy with what done in Section \ref{par:NoetherAnalysis})
one obtains by spatial integration of the $\bar{J}_{0 A}$
some conserved quantities $\bar{Q}_A$ which are themselves simply related to
the $\hat{Q}_A$ derived in Section \ref{par:NoetherAnalysis}:
\begin{equation}\label{Qbarra}
\!\left(\begin{array}{c}
\bar{Q}_0\\
\bar{Q}_i\\
\bar{Q}_4\\
\end{array}{}\right)
\!=\!-\frac{1}{2}\int  d^4k \left|\tilde{\Phi}(k)\right|^2
\left(\begin{array}{c}
\hat{k}_0+\hat{k}_4\\
\sqrt{2}\,\hat{k}_i\\
\hat{k}_0-\hat{k}_4\\
\end{array}{}\right) \, \frac{(-2\tilde{k_0}e^{\frac{\lambda}{2}k_0}+\lambda
m^2)}{|-2\tilde{k_0}e^{\frac{\lambda}{2}k_0}+\lambda
m^2|}\,\delta(C_{\lambda}(k)-m^2)\!=\!\!
\left(\begin{array}{c}
\hat{Q}_0+\hat{Q}_4\\
\sqrt{2}\,\hat{Q}_i\\
\hat{Q}_0-\hat{Q}_4\\
\end{array}{}\right)\,.
\end{equation}
In spite of the simplicity of the formula giving $\bar{Q}_0$ from $\hat{Q}_0$ and $\hat{Q}_4$,
from a conceptual perspective $\bar{Q}_0$ might turn out to be a valuable tool,
since it is the conserved charge associated with the transformation parameter $\bar{\epsilon}_0$,
and therefore (in light of the fact that
in $\bar{d}f$ we have $\bar{\epsilon}_0$ multiplying $\bar{\mathcal{D}}_0$,
which is a plausible time-translation generator)
is a plausible candidate for the energy charge.

\section{Some remarks on relativistic structures}\label{par:Remarks}
As mentioned in the Introduction a primary source of interest
in $\kappa$-Minkowski comes from the intuition that theories in this
spacetime should be subject to a new type of spacetime symmetries.
While the fact that we are finally able to perform some Noether
analyses should prove useful for clarifying this possibility, we
feel that at present nothing definite can be said. A key point is
that, both using the 4D differential calculus~\cite{kappanoether}
and the 5D differential calculus, one obtains charges that are time
independent but for which it is not obvious how one should introduce
a prescription for measurement. This is particularly true for the
analysis based on the standard basis for the 5D calculus: the analysis itself suggests that
none of the transformation parameters is a time-translation parameter,
and consequently none of the charges can be viewed as a charge
conserved under time-translation symmetry.
The change of basis for the 5D differential calculus which we proposed in the
previous section does lead to a ``candidate time-translation-symmetry charge"
which at least is plausible, but several logical-consistency checks (starting
indeed by asking ``how should one measure $\bar{Q}_0$?")
should be performed before any definite claim.

This point is also relevant for the ``equivalence between a
free $\kappa$-Minkowski field theory and a
free relativistically invariant (non-local) field theory on
classical Minkowski space-time" proposed in Ref.~\cite{kowafreidel}.
We believe that at present such an equivalence simply cannot
be established. This can be seen in many ways (comparing structures which emerged
in our analysis to structures available in ordinary commutative theories),
and perhaps most clearly by considering that on the commutative-theory side of
the ``equivalence" proposed in Ref.~\cite{kowafreidel} the energy
observable is readily available while on the non-commutative-theory side
(since Ref.~\cite{kowafreidel} makes reference to the standard basis for
the 5D calculus)
there is the mentioned issue concerning the energy observable.

Together with the energy-observable issues several other structures must be
better understood before formulating any definite statement on relativistic properties.
One (of possibly many) point to consider concerns the difference between
the factor $\delta(C_{\lambda}(k)-m^2)$ in our formulas and the
factor $\delta(k_0^2-\vec{k}^2-m^2)$ in the corresponding formulas found
for theories in classical commutative Minkowski spacetime.
It is perhaps noteworthy that
\begin{equation*}
\delta(C_{\lambda}(k)-m^2)=\delta\left((\frac{2}{\lambda}\sinh{\frac{\lambda
k_0}{2}})^2-|\vec{k}|^2e^{\lambda k_0}-m^2\right)=
\end{equation*}
\begin{equation}
=\frac{1}{2\sqrt{m^2+|\vec{k}|^2+\lambda^2m^4/4}}(\delta(k_0
-k^+_0)+\delta(k_0 -k^-_0)) ~,
\end{equation}
where
\begin{equation}
k^+_0=\frac{1}{\lambda}\ln\left(\frac{1+(\lambda m)^2/2+\lambda
\sqrt{m^2+|\vec{k}|^2+\lambda^2m^4/4}}{1-(\lambda
|\vec{k}|)^2}\right)
\label{k0+}
\end{equation}
\begin{equation}
k^-_0=\frac{1}{\lambda}\ln\left(\frac{1+(\lambda m)^2/2-\lambda
\sqrt{m^2+|\vec{k}|^2+\lambda^2m^4/4}}{1-(\lambda
|\vec{k}|)^2}\right) ~,
\label{k0-}
\end{equation}
and $k^+_0$ is real only for $|\vec{k}|<{1}/{\lambda}$.
But here a complex $k^+_0$ may well
be admissible\footnote{Note however that there appears to be no obstruction
for implementing the restriction $|\vec{k}|<1/\lambda$ on the Fourier
parameters. In fact, in a recent proposal of description of quantum fields
in $\kappa$-Minkowski~\cite{ArMa} this restriction was analyzed in some detail,
finding that, upon adopting a suitable inner-product
for the Hilbert space, it
does not constitute a source of incompleteness in the construction
of the Hilbert space. Instead by allowing complex values of $k^+_0$, and
therefore $|\vec{k}|>{1}/{\lambda}$, Ref.~\cite{ArMa} found that the
inner product was no longer guaranteed to be positive definite.}, since in our
theory the criterion one should enforce, which is the one
of a real and positive energy charge, is at present not
manageable (we do not know how to describe the energy observable,
so we cannot test its reality and positivity).

Another key indicator of the relativistic structure of a theory is
the energy-momentum dispersion relation, and of course also this
indicator will not be available until a robust description of the
energy observable is discovered. We thought that it might be worth looking for
possible striking invariant combinations of the charges we obtained,
but we found nothing noteworthy. One could
attempt to identify such
a combination of charges by probing the interconnection between the
charges through a regularized plane-wave $\Phi^{p.w.}({x})$
solution of the equation of motion, of the following form:
\begin{equation}
\Phi^{p.w.}({x})=\frac{1}{(2V\sqrt{m^2+|{k}|^2
+\lambda^2m^4/4})^{\frac{1}{2}}}e^{i\vec{k}\cdot\vec{{x}}}e^{-ik^+_0\, {x}_0},
\end{equation}
where $V$ is a spatial-volume normalization factor and $k^+_0$ is
related to $|\vec{k}|$ by (\ref{k0+}).

Our results attribute to this field the charges
\begin{eqnarray}
\hat{Q}^{p.w.}_0\!\!&=& \hat{k}_0 \big|_{k^+_0, \vec{k}}\,,\nonumber\\
\hat{Q}^{p.w.}_i\!\!&=&\hat{k}_i \big|_{k^+_0, \vec{k}}\,,\nonumber\\
\hat{Q}^{p.w.}_4\!\!&=& \hat{k}_4 \big|_{k^+_0, \vec{k}} \,.\label{cacom}
\end{eqnarray}
It is perhaps intriguing that
\begin{eqnarray}
(\hat{Q}^{p.w.}_0)^2-(\hat{Q}^{p.w.}_i)^2 &=& m^2+(\hat{Q}^{p.w.}_4)^2
=m^2\left(1+\frac{\lambda^2m^2}{4}\right)
\end{eqnarray}
but this should be analyzed taking into consideration the fact that,
in light of the observations we reported above on time translations, $\hat{Q}^{p.w.}_0$
clearly cannot be the energy carried by our regularized plane wave.

In light of the observations we reported in the previous section
one might consider contemplating a role for the
combination $\hat{Q}^{p.w.}_0+\hat{Q}^{p.w.}_4$ (which gives
a ``$\bar{Q}_0^{p.w.}$" for our regularized plane wave),
but we could not find any good use for it. For
example, we find that
\begin{equation}
\left(\hat{Q}^{p.w.}_0+\hat{Q}^{p.w.}_4\right)^2
-\left(\hat{Q}_i\right)^2=\left(\frac{e^{\lambda
k^+_0}-1}{\lambda}\right)^2-\left({k_i}e^{\lambda k^+_0}\right)^2=
\left(\lambda \hat{Q}^{p.w.}_0+ \lambda \hat{Q}^{p.w.}_4 +1\right)m^2 ~.
\label{eq:Dispersion}
\end{equation}

\section{Comparison with a recent related analysis}\label{par:Comparison}
While we were in the final stages of
preparation of this manuscript we became aware of the very recent
 Ref.~\cite{kowafreidel2}, by the same authors of Ref.~\cite{kowafreidel},
which considers the same framework we analyzed here (and they already
analyzed in Ref.~\cite{kowafreidel}).
While it is probably fair to say that there is absolutely no overlap
between the analysis we are here reporting and the one in
Ref.~\cite{kowafreidel},
there is a correspondence between at least some
points of our analysis and some corresponding points of
the analysis reported in Ref.~\cite{kowafreidel2}.
But a large number of crucial differences remain, and we hope to contribute
to future further studies of this framework by stressing these differences,
while acknowledging the points in common.

One first point of contact is that, while the analysis in
Ref.~\cite{kowafreidel} was only rather vaguely inspired by the 5D-calculus setup
(which never explicitly appeared in the analysis),
 Ref.~\cite{kowafreidel2} uses explicitly the 5D calculus to construct a $df$,
 with just the same perspective and the same results we reported in Section II.

We should also mention that, while Ref.~\cite{kowafreidel} only
obtained 4 conserved charges from the 5D-calculus setup,
Ref.~\cite{kowafreidel2} reports 5 conserved charges from the
5D-calculus setup. This is qualitatively consistent with our
identification of 5 conserved charges. However, at the quantitative
level (comparing the structure of the charges rather than just their
abundance) there are significant differences between our results and
the ones of Ref.~~\cite{kowafreidel2} (and with the ones of
Ref.~\cite{kowafreidel}), and it is not hard to understand how these
differences have emerged. In fact, while our Noether analysis
constructively led us to a ``conservation equation" of the form
${\mathcal{D}}_{0}\,J^{0}_A+\hat{P}_{i}\,J^{i}_A=0$,
Ref.~\cite{kowafreidel2} somehow arrives at a conservation equation
of the type\footnote{We have chosen, as a way to render the
discussion clearer for our readers, to use our notation consistently
throughout, even when reporting equations from
Ref.~\cite{kowafreidel2}. For a direct comparison with what written
in Ref.~\cite{kowafreidel2} it should be noticed that our
$\hat{P}_{\mu}$ is denoted by $\hat{\partial}_{\mu}$ in
Ref.~\cite{kowafreidel2}, while for our ${\mathcal{D}}_{0}$
there is no dedicated symbol in Ref.~\cite{kowafreidel2} (our
${\mathcal{D}}_{0}$ would be described within the conventions
adopted in Ref.~\cite{kowafreidel2} as
$\hat{\partial}_{0}+\hat{\partial}_{4}$).}
$\hat{P}_{\mu}\,J^{\mu}_A=0$.

We have been unable to identify the assumption or choice which could
have caused the analysis of Ref.~\cite{kowafreidel2} to end up with
a $\hat{P}_{\mu}\,J^{\mu}_A=0$ pseudo-conservation equation.
However, we are confident that any recipe
of Noether analysis leading to ``would-be-conservation laws" of
the form $\hat{P}_{\mu}\,J^{\mu}_A=0$ should be rejected since, in light
of the mentioned inadequacy of $\hat{P}_{0}$ to describe time derivatives,
these are simply unacceptable as
conservation equations.

Besides this crucial difference there are clearly many other
differences between the two analyses, but
the exercise to fully translate the analysis reported in Ref.~\cite{kowafreidel2}
into formulas that are meaningful within our formulation of the problem
is rather complex. One obstruction is caused by the fact that,
while the equation of motion for scalar fields that we considered is the
most studied such equation in the $\kappa$-Minkowski literature,
 Ref.~\cite{kowafreidel2} performs a symmetry analysis for a somewhat
 different, less known, equation of motion for scalar fields,
 which had been previously proposed by one of the authors.
And a further difficulty is introduced by
the fact that some of the formulas in Ref.~\cite{kowafreidel2}
appear to be intended for quantum fields (although no Hilbert-space construction
is offered), since they
are described in terms of operators counting the number of particles; here instead
we focused on the case of a Noether analysis of classical fields in our noncommutative spacetime.

\section{Summary and outlook}\label{par:ClosingRemarks}
While in Ref.~\cite{kappanoether}, where some of us reported a
first example of successful Noether analysis of Hopf-algebra
spacetime symmetries, the description of translation
transformations in 4D $\kappa$-Minkowski spacetime was based on
the properties of a 4D differential calculus, more recently, in
Ref.~\cite{kowafreidel}, it had been argued that there should also be a
description of translations in 4D $\kappa$-Minkowski spacetime
inspired by a 5D differential calculus. And while the derivation
reported in Ref.~\cite{kappanoether}
 required some rather tedious manipulations of
noncommutative functions and operator coproducts,
the Noether analysis reported in Ref.~\cite{kowafreidel}
relied on a proposed
equivalence between a free $\kappa$-Minkowski field theory and a
free relativistically invariant (non-local) field theory on
classical Minkowski space-time. We here exposed some
limitations to the applicability of the proposed equivalence of
theories, which in particular, as we showed, associates the operator of time
derivation on the commutative-theory side to an operator which
does not even vanish on time-independent functions on the noncommutative-theory side.

However we also provided here further evidence
in support of the
possibility that the 5D-calculus-based ``translation transformations"
can indeed be implemented as symmetries of theories in $\kappa$-Minkowski.
Our analysis performed
directly within the noncommutative theory also allowed us to
investigate explicitly the properties of the 5 ``would-be
currents" that one naturally ends up considering when working with
the 5D calculus.

The fact that the techniques developed by some of us in
Ref.~\cite{kappanoether} for a 4D-calculus-based description of
translations were here successful, without any need of adaptation,
in dealing with the very different 5D-calculus-based description of
translations certainly encourages the hope that these techniques may
be robust enough to deal with any kind of Hopf-algebra spacetime
symmetries. We believe that a particularly striking indicator of the
robustness of these techniques is provided by the fact that they
automatically fixed an apparent problem of the standard basis for the 5D
differential calculus, which leads to
 a $0$-label generator acting in a way that would be inacceptable
 for  a time-translation generator. Our approach constructively led us to
current-conservation-like equations
written in terms of the operator ${\mathcal{D}}_{0}$ which
instead is a plausible candidate for the generation of time translations.

Concerning the puzzling apparent availability of different descriptions
of translations transformations in $\kappa$-Minkowski our analysis did not
lead to a definite answer, but, just because we showed that the description
of the energy observable must be rather ``tricky" within the 5D-calculus-based
setup, it is legitimate to be hopeful: it is plausible that once we will have a
robust understanding of the energy-momentum observables both in the
4D-calculus-based description and in the 5D-calculus-based description these two
descriptions of translation transformations may turn out to be equivalent.
The change of basis which we proposed in Section V may well turn out to be
useful for this task.

The challenge of a proper identification of energy-momentum observables
is also a necessary first step toward addressing the most significant
issue here of interest, which concerns the fate of physical/observable
aspects of spacetime symetries in noncommutative geometry.
For this, besides the energy-momentum observables, one would also need to address
other issues, some of which were preliminarily
considered in our Section~\ref{par:Remarks}.

\section*{Acknowledgments}
We thank Pierre Martinetti and Ruggero Altair Tacchi for valuable
comments, especially  during the early stages of this analysis.

 \end{document}